\documentclass[aps,floatfix,footinbib, preprint,superscriptaddress,amsmath,amssymb,amsfonts]{revtex4-1}

\usepackage[T1]{fontenc}
\usepackage{lmodern} 
\usepackage{color}
\usepackage{bbold}
\usepackage{dsfont}
\usepackage{graphicx}
\usepackage[caption=false]{subfig}
\usepackage[colorlinks]{hyperref}
\usepackage[bottom]{footmisc}
\usepackage{enumerate}
\usepackage{float}

\usepackage{empheq}
\usepackage{tikz}
\usepackage{fancyhdr}
\usepackage[normalem]{ulem}
\DeclareMathAlphabet\mbc{OMS}{cmsy}{b}{n}
\usepackage{balance}

\begin{document}

\title{Torsional optomechanical cooling of a nanofiber}

\author{Dianqiang Su}
\affiliation{State Key Laboratory of Quantum Optics and Quantum Optics Devices, Institute of Laser Spectroscopy, Shanxi University, Taiyuan 030006, People's Republic of China.}
\affiliation{Collaborative Innovation Center of Extreme Optics, Shanxi University, Taiyuan 030006, People's Republic of China.}

\author{Pablo Solano}
\affiliation{Departamento de F\'isica, Facultad de Ciencias F\'isicas y Matem\'aticas, Universidad de Concepci\'on, Concepci\'on, Chile.}

\author{Jeffrey D. Wack}
\affiliation{Joint Quantum Institute, Department of Physics and NIST, University of Maryland, College Park, MD 20742, USA.}

\author{Luis A. Orozco}
\affiliation{Joint Quantum Institute, Department of Physics and NIST, University of Maryland, College Park, MD 20742, USA.}

\author{Yanting Zhao }
\email{zhaoyt@sxu.edu.cn}
\affiliation{State Key Laboratory of Quantum Optics and Quantum Optics Devices, Institute of Laser Spectroscopy, Shanxi University, Taiyuan 030006, People's Republic of China.}
\affiliation{Collaborative Innovation Center of Extreme Optics, Shanxi University, Taiyuan 030006, People's Republic of China.}

\begin{abstract}
We demonstrate the optomechanical cooling of a tapered optical nanofiber by coupling the polarization of light to the mechanical angular momentum of the system. The coupling is enabled by birefringence in the fiber and does not make use of an optical resonator. We find evidence for cooling in the distribution of thermally driven amplitude fluctuations and the noise spectrum of the torsional modes. Our proof-of-principle demonstration shows cavity-less cooling of the torsional degree of freedom of a macroscopically extended nanofiber.
\end{abstract}

\maketitle

\section{Introduction}

Optomechanical cooling is the reduction of thermally induced noise in the displacement of a mechanical object by its controlled interaction with light. So far, experimental realizations of it rely on the transfer of linear momentum between light and the object~\cite{Chan2011,Delic2020,Otterstrom2018}. However, the transfer of angular momentum, by means of the light orbital angular momentum~\cite{He1995,Bhattacharya2007,Bhattacharya2015} or polarization~\cite{friese98,Tong2010,arita13,He2016,shi16,hoang16,Kuhn17Optica,kuhn17,reimann18,Delord2020}, has never been used to reduce the mechanical fluctuations of a torsional oscillator. Here we demonstrate in a proof-of-principle experiment the reduction of such mechanical fluctuations by transferring angular momentum between the light and the object.

Linearly polarized light can exert a torque on a birefringent object when passing through it \cite{Pan2021}. This phenomenon was revealed by Poynting~\cite{poynting09} and beautifully demonstrated in the pioneer experiments of Beth~\cite{beth35,beth36} and Holbourn~\cite{holbourn36}. A next milestone for controlling movable massive objects by angular momentum exchange with light is cooling. Recent experiments have observed light polarization optomechanical coupling to this end with levitated nanodumbbells \cite{Bang2020}. To enable significant optomechanical effects one benefits from lightweight objects and large optical intensities, two properties present in small optical waveguides which highly confine the spatial modes of propagating light. In particular, torsional optomechanical coupling has already been observed using the torsional modes of optical nanofiber (ONF) waveguides~\cite{fenton18}. 

An ONF waveguide is a macroscopic object along one direction and nanometric in the transverse dimensions, confining the propagating light to extremely large intensities~\cite{solano17d} and enhancing optomechanical effects. The nanofiber resides in a vacuum chamber fixed at two points in the unmodified region of the optical fiber. It generally has vibrational, compressional, and torsional modes~\cite{wuttke14,Hummer2019}. Here we focus on the torsional ones, with angular momentum pointing along the nanofiber. The thermally induced internal molecular vibrations of the ONF set the torsional modes in motion \cite{wuttke13}. The first torsional mode has nodes at each end before the tapered region, and an antinode at the center of the ONF (see Ref.~\cite{wuttke14} for more information on the shape of the modes). Such mechanical motion lies between macroscopic and microscopic scales, since the torsional mode is present along the entire nanofiber which has an aspect ratio of $10^{6}$ between its transverse and longitudinal dimensions.

The optically induced forces change the effective torsional spring constant and, more importantly, the torsional damping, allowing for cooling. The changes in the rotational amplitude fluctuations and the width of its noise spectrum indicate a temperature reduction. The method does not rely on an optical resonance, but only on the intensity and polarization of the propagating light. 

\section{Theoretical Model}
We model the effects of light polarization by considering a section of the nanofiber as a birefringent disk. Its tensor susceptibility allows light to induce a macroscopic polarization with a component perpendicular to the incident electric field. The light and medium polarizations are parallel to each other only for two particular axes, and we call the one with smaller index of refraction the optical axis of the disk. The torque applied by the electric field of the light to a polarizable medium is $ \Vec{\tau}=\int dA(\Vec{P} \times \Vec{E})$,  where $\Vec{P}$ is the polarization of the medium, $\Vec{E}$ is the electric field in the medium and the integral extends over the interaction area. 
Such torque
is a function of the angle between the electric field and the optical axis $\Delta\theta$ as~\cite{beth36,friese98}:
\begin{equation}
    \tau=\tau_{0}\sin\left(2\Delta\theta\right),
 \label{torque}   
 \end{equation}
with
 \begin{equation}
    \tau_0=\frac{c\epsilon }{2 \omega_{\text{L}}}|E|^2\pi r_0^2\sin{\eta},
\label{torque0}
\end{equation}
where $\epsilon$ is the electric permitivity related to the effective index of refraction of the ONF, $c$ is the speed of light, $\omega_{\text{L}}$ is the laser frequency, $E$ is the amplitude of the electric field of the drive, $r_0$ is the ONF radius, and $\eta=k d(n_{\parallel}-n_{\perp})$ with $k$ the wavenumber, $d$ the dielectric thickness, and $n_{\parallel}$ and $n_{\perp}$ the indices of refraction parallel and orthogonal to the optical axis respectively. $\Delta \theta=\theta_{\rm{F}}-\theta_{\rm{L}}$, where $\theta_{\rm{L}}$ is the polarization angle of the driving laser and $\theta_{\rm{F}}$ is the optical axis of the disk, which we use to characterize the nanofiber rotation. We assume a homogeneous light intensity distribution across the ONF cross-section. 

Considering the origin of the angular coordinate at the
equilibrium point of the nanofiber rotation angle $\theta_{\rm{F}}$,
we can expand to first order $\sin(2\Delta\theta(t))\approx2\theta_{\rm{F}}(t)\cos(2\theta_{\rm{L}}(t))-\sin(2\theta_{\rm{L}}(t))$.
The laser angle $\theta_{\rm{L}}(t)$ is arbitrary and determines the sign
of the torque applied to the ONF. The total torque can be then decomposed
into two contributions: a spring force (proportional to the ONF
angle) and an offset force, as
\begin{equation}
\tau_{\rm{ext}}(t)=\tau_{o}(t)+\theta_{\rm{F}}(t)\kappa_{\rm{L}}(t),
\label{eq:torque2}
\end{equation}
where $\tau_{o}(t)=-\tau_{0}\sin(2\theta_{\rm{L}}(t))$ and $\kappa_{\rm{L}}(t)=2\tau_{0}\cos(2\theta_{\rm{L}}(t))$. In optimum cool/heating conditions $2\theta_{\rm{L}}=n\pi$, and the laser
angle will only perturbatively change due to the ONF back-action, allowing to approximate $\tau_{o}(t)=-2\tau_{0}\theta_{\rm{L}}(t)$ and $\kappa_{\rm{L}}=2\tau_{0}$. 

A rotational harmonic oscillator describes the torsional motion of the disk ($\theta_\text{F}\ll1$) with a driving force of the form of Eq. (\ref{eq:torque2}). The equation of motion, in the Langevin form, for the generalized coordinate (angle) of a rotating disk with the torque along the light propagation axis is:
\begin{equation}
I \ddot{\theta}_{\text{F}}+\gamma \dot{\theta}_{\text{F}}+\kappa \theta_{\text{F}}=\tau_{\rm{th}}+\tau_{\rm{ext}},
\label{eqmotion}
\end{equation}
where $I$ is the moment of inertia, $\gamma$ is the damping coefficient, $\tau_{\rm{th}}$ is the thermally induced torque, and $\tau_{\rm{ext}}$ is the torque induced by the driving light. The thermally induced torque $\tau_{\rm{th}}$ has a white power spectral density:
\begin{equation}
S_{\tau_{\rm{th}}}=4k_BT\gamma,
\label{SD_T}
\end{equation}
where $T$ is the temperature of the system and $k_B$ is the Boltzmann constant.

Modeling the ONF as a simple birefringent disk does not consider the mechanical effects along its entire length. For an extended mechanical system the optically induced torque, from Eq.~(\ref{eq:torque2}), generally includes two effects: an instantaneous response of the system to light, which modifies the torsional constant $\kappa$; and a delayed response, due to local mechanical perturbations propagating through the extended system, which modifies the damping coefficient $\gamma$ \cite{Metzger2004}.
When torque is applied
at a given position, sound waves propagate along the nanofiber affecting its torsional modes, leading to a delayed response to local perturbations. We model this behaviour with an arbitrary response function $h(t)$ of the system, rewriting the equation of motion Eq.~(\ref{eqmotion}) for the rotation angle of a cross-section of the ONF as
\begin{equation}
I\ddot{\theta_{\rm{F}}}(t)+\gamma\dot{\theta_{\rm{F}}}(t)+\kappa\theta_{\rm{F}}(t)=\tau_{\rm{th}}(t)+\int_{0}^{t}\dot{\tau}_{\rm{ext}}(t)h(t-t')dt'.
\label{eq:motion2}
\end{equation}

We can solve the differential equation by taking its Fourier transform, since we are interested in the long term response of the system, meaning $t\gg1$, approximating the time integral to a convolution. The Fourier transform of the response function will generally have a real and imaginary part, namely $h_\omega=h_\omega^{\rm{Re}}+i h_\omega^{\rm{Im}}$, which will be related to energy dissipation and dispersion respectively. Because the external force is an oscillatory function, the Fourier transform of the time derivative of Eq. (\ref{eq:torque2}) can be written as
$\mathcal{F}\left\{ \dot{\tau}_{\rm{ext}}(t)\right\}=i\omega\tau_{\rm{ext},\omega}=i\omega(\tau_{o,\omega}+\kappa_{\rm{L}}\theta_{\rm{F},\omega}).$ Collecting the terms for the equation of motion Eq.~(\ref{eq:motion2}) in the Fourier space we can find the amplitude of the rotation angle of a given cross-section of the ONF as

\begin{equation}
\theta_{\rm{F},\omega}=\left(\frac{\tau_{\rm{th},\omega}+i\omega h_{\omega}\tau_{o,\omega}}{\kappa}\right)\left(\frac{\omega_{0}^{2}}{\omega_{\text{eff}}^{2}-\omega^{2}+i\omega\Gamma_{\text{eff}}}\right),
\label{SMeqSpectrum}
\end{equation}
where we define the bare resonance frequency $\omega_{0}^{2}=\kappa/I$, the damping rate $\Gamma=\gamma/I$, and where the resonance frequency and damping have been both modified by the external torque and the response function of the system as
\begin{eqnarray}
\omega_{\text{eff}}^{2}=&\omega_{0}^{2}\left[1+\frac{\kappa_{\rm{L}}}{\kappa}\omega_{0} h_{\omega}^{\rm{Im}}\right],\label{SM.Eq.oemga}\\
\Gamma_{\text{eff}}=&\Gamma\left[1-Q\frac{\kappa_{\rm{L}}}{\kappa}\omega_0 h_{\omega}^{\rm{Re}}\right], \label{SM.Eq.Gamma}
\end{eqnarray}
with $Q$ the quality factor of the torsional
mode and $\kappa_{\text{L}}$ the light-induced effective spring constant from Eq. (\ref{eq:torque2}). The effective damping constant can be written as the sum of the bare damping and a damping coming from the external drive, $\Gamma_{\rm{eff}}=\Gamma+\Gamma_{\rm{L}}$. Eq. (\ref{SMeqSpectrum}) shows that the amplitude of the rotation is set by the thermally induced torque
plus some constant offset force from the light. Such offset force approaches zero for optimal cooling conditions ($\theta_{\rm{L}}(t)=0$).

The thermal torsional power
in a frequency window $\delta f=\delta\omega/2\pi$ is
|$\tau_{\rm{th},\omega}|^{2}=4k_{B}T\gamma\delta f$ \cite{saulson90}.
This allows to obtain an expression for the power spectral density (PSD) of the thermal fluctuations of the mode around the steady state of Eq. (\ref{eq:motion2}). The PSD for a small perturbation of the rotation angle $\theta_{\text{F}}$, around a fixed stable value, in the vicinity of one of its torsional resonances, is
\begin{equation}
|\theta_{\rm{F},\omega}|^{2}=S_{\theta_{\text{F}}}(\omega)\delta\omega/2\pi=\left(\frac{4k_{B}T}{\kappa}\right)\left(\frac{\omega_{0}^{2}\Gamma}{\left(\omega_{\text{eff}}^{2}-\omega^{2}\right)^{2}+\omega^{2}\Gamma_{\text{eff}}^{2}}\right)\delta\omega/2\pi.
\label{spdentheta}
\end{equation}

The noise in Eq. (\ref{spdentheta}) can increase or decrease depending on the polarization angle of a guided field with respect to the optical axis of the ONF waveguide. Such optomechanical effect is encoded in the amplitude and sign of $\omega_{\rm{eff}}$, $\kappa_{\rm{L}}$ and $\Gamma_{\rm{eff}}$. When light adds optical damping $\Gamma_{\rm{L}}$, without simultaneously adding a Langevin force, it reduces the roto-mechanical fluctuations, meaning cooling. Upon integration of the PSD in Eq.~(\ref{spdentheta}) over frequency we get
\begin{equation}
\frac{1}{2}\kappa\langle|\theta_{\rm{F}}|^{2}\rangle=\frac{1}{2}k_{B}T\left(\frac{\omega_{0}^{2}\Gamma}{\omega_{\text{eff}}^{2}\Gamma_{\text{eff}}}\right).
\label{PE-KE}
\end{equation}
This expression relates the torsional potential energy to the
thermal energy, and it is consistent with the equipartition theorem if we define
the effective temperature
\begin{equation}
T_{\text{eff}}= T\left(\frac{\Gamma}{\Gamma_{\text{eff}}}\right),
\label{SM.Eq.T}
\end{equation}
where we assumed $\omega_0\approx\omega_{\rm{eff}}$. The effective temperature of the torsional mode is then the ratio of the half width at half maximum (HWHM) of the spectral density with drive, $\Gamma_{\rm{eff}}=\Gamma+\Gamma_{\rm{L}},$ and without drive, $\Gamma$.

Increasing damping could, in principle, come from dissipation induced by extra Langevin forces, so a spectral broadening is not an unequivocal signal of cooling. In order to demonstrate cooling, the overall integrated PSD should decrease. An equivalent unequivocal proof of cooling comes from the narrowing of the distribution of amplitude fluctuations, which in thermal equilibrium follows a Maxwell-Boltzmann (MB) distribution. Given Eq. (\ref{eq:torque2}), the potential energy $U(\theta_{\rm{F}})=\frac{1}{2}\kappa \theta_{\rm{F}}^2$ allows to obtain the probability of finding the system with energy between $U$ and $U+dU$ as $p(U)dU\propto e^{-U/k_{\rm{B}}T}dU$. Then, the probability $p(\theta_{\rm{F}})$ of finding the system between the rotation angles $\theta_{\rm{F}}$ and $\theta_{\rm{F}}+d\theta_{\rm{F}}$ is related to $p(U)$ as $p(U)dU=p(\theta_{\rm{F}})d\theta_{\rm{F}}$, leading to the ONF angle probability distribution
\begin{equation}
    p(\theta_{\rm{F}})d\theta_{\rm{F}}=\frac{\kappa \theta_{\rm_{F}}}{k_{\rm{B}} T_{\rm{eff}}}e^{-\kappa \theta_{\rm{F}}^2/2 k_{\rm{B}} T_{\rm{eff}}}d\theta_{\rm{F}}.
\label{MB} 
\end{equation}
One can first make a test that the MB distribution is appropriate by calculating the ratio of $\langle\theta^2_{\rm{F}}\rangle/\langle\theta_{\rm{F}}\rangle^2$ and find numbers consistent with the expected result of $4/\pi$. We measure the a voltage output $V$ proportional to the polarization rotation and assume that is linearly proportional to fiber angle such that $\theta_{F}=\xi V$. We use the changes in the MB distribution and the PSD extracted from $V$ to characterize our results.

We see from Eqs. (\ref{SM.Eq.Gamma}) and (\ref{SM.Eq.T}) that there will be optomechanical cooling when the Fourier transform of the response function, $\mathcal{F}\left\{ h(t)\Theta(t)\right\}$, has a non-zero real component, $h_{\omega}^{\rm{Re}}$. In general, the response function of the system is difficult to determine. The simplest function that represent a delayed response of the system is an exponential delay. We will consider this $h(t)=1-e^{-t/\mathcal{T}}$, where $\mathcal{T}$ is a characteristic time response. The time scale for the delayed response will be set by the speed of sound propagating along the nanofiber, typically of the order of $\mathcal{T}\sim 1/\omega_0$. Then, near the resonance frequency we get
\begin{eqnarray}
\omega_{\text{eff}}^{2}\sim& \omega_{0}^{2}\left[1-\frac{\kappa_{\rm{L}}}{2\kappa}\right],\label{SM.Eq.omega2}\\
\Gamma_{\text{eff}}\sim& \Gamma\left[1+Q\frac{\kappa_{\rm{L}}}{2\kappa}\right]. \label{SM.Eq.Gamma2}
\end{eqnarray}
With this approximations Eqs. (\ref{SM.Eq.T}) and (\ref{SM.Eq.Gamma2}) predict a linear dependence of the spectral width and the effective temperature with the drive power, since $\kappa_{\rm{L}} \approx2\tau_0\propto |E|^2$. 

Optomechanical cooling mechanisms rely on feedback, either active, passive or a combination of both, as it is often the case in cavity assisted cooling. In our case, the feedback is passive, and it comes from the delayed mechanical response of the ONF do to the finite speed at which mechanical waves propagate through it. Our simplified model describes a waveplate-like cylindrical section within the ONF, but a description of the full nanofiber dynamics is more complicated. Moreover, the ONF optical axis $\theta_{\rm{F}}$ is unknown a priori and can come from a combination of three different sources of birefringence, as Eq.(\ref{torque0}) has to be different from zero: (i) Torsional stress-induced birefringence. Given the reflection symmetry of the torsional modes relative to a cross-sectional plane at the center of the ONF, there should not be a net light polarization rotation due to the stress-induced by the torsional modes~\cite{wuttke14}. However, slight asymmetries sometimes seen in the tapered regions of the ONF can lead to net birefringence. (ii) Intrinsic birefringence. Stress or imperfections unintendedly produced during the fabrication can lead to a permanent birefringence. (iii) Light-induced  through a Kerr effect. The light used in our detection scheme might also be a source of birefringence.

A more detailed description of the optomechanical effect of the light polarization on the torsional modes of an extended mechanical system goes beyond the scope of this paper.

\section{Experimental setup}
\begin{figure}
\begin{center}
\includegraphics[width=7cm]{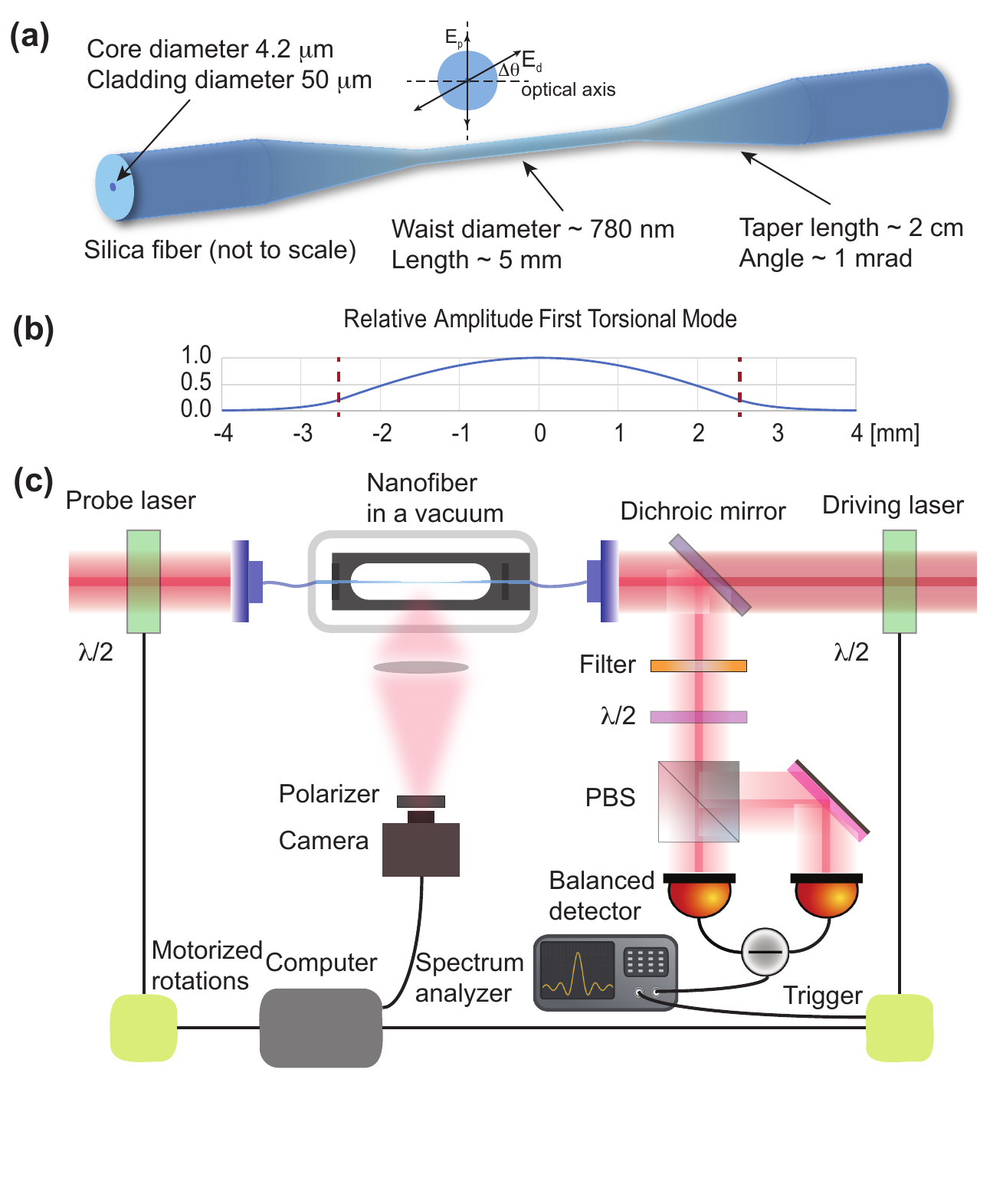}
\caption{ \textbf{(a)} Schematic of the ONF with two effective polarization axis associated with an ordinary and an extraordinary indices of refraction, aligned with the optical axis, but at an angle $\Delta\theta$ with the input light polarization. The fiber is clamped (not shown) in the unmodified section. {\textbf{(b)}} Calculated relative amplitude of the first torsional mode. The red dashed lines indicate where the waist ends and the torsion extends into the tapered region. {\textbf{(c)}} Schematic of the apparatus. Probe and drive laser beams of 852 nm and 780 nm wavelength respectively counter-propagate with independent polarization control by motorized half-waveplates. The signal output goes to a half-waveplate to set the detection basis and then to a polarizing beam splitter (PBS) that separates the light into two polarizations components, falling into a balanced photodiode pair.}

\label{apparatus}
\end{center}
\end{figure}

Our apparatus, schematically drawn in Fig.~\ref{apparatus}, operates at room temperature. The fiber is produced with the heat and pull method \cite{hoffman14}, using a stripped down commercial optical fiber Fibercore SM1500 with no specified birefringence. The 780 nm diameter, 5 mm waist length, and 1 mrad taper of the nanofiber allows the propagation of the fundamental HE$_{11}$ mode and some higher order modes \cite{ravets13} at the experimental wavelengths. The nanofiber resides in a vacuum chamber. Since ONFs do not preserve the polarization of propagating light, we analyze the Rayleigh scattering polarization from a side window. We tune the input polarization to guarantee mostly linearly polarized light at the nanofiber waist \cite{Joos2019}, which maximizes the optically-induced torque in all experimental configurations. The probe and drive lasers are Toptica DLS pro. The transmission of the ONF at the drive and probe wavelengths (780 nm and 852 nm) is better than  84\% including the coupling efficiency into and out of the ONF.

The ONF has its first torsional mode resonance $\nu_{T1}$ at about 190 kHz with a full width at half maxima (FWHM) of 3 Hz, corresponding to a $Q\sim 0.6\times 10^5$. Other nanofibers fabricated with the same specification show a similar $Q$ and resonant frequencies differing by less than 400 Hz. The estimated waist effective length is 9.9 mm, given the gradual exponential taper of the ONF. This corresponds to frequencies for the first compressional and flexural modes of around 290 kHz and 18 Hz respectively \cite{Hummer2019}. Since they are away from the frequency of the first torsional mode and have negligible coupling to light polarization, we can be sure of just manipulating the torsional motion. The calculated relative amplitude of the first torsional mode is shown in Fig.~\ref{apparatus}(b).

We use a linearly polarized weak probe at 852 nm ($\approx50~\mu$W), propagating through the ONF as a proxy to measure the nanofiber rotation. We monitor both linear components of its polarization with a Balanced Photo Diode (BPD), Thorlabs PDB450A, to filter common noise (see Fig.~\ref{apparatus}(c)) and obtain a voltage proporional to the polarization rotation. We send the BPD main output to a Rohde \& Schwarz FSVA13 spectrum analyzer, that can work as a narrow-band filter for a time series measurement or directly produce the amplitude spectrum. We continuously monitor the BPD independent outputs to ensure balanced operation. 

The photocurrent is linearly proportional to the ONF rotation angle. We record the thermal noise of the rotation angle around the resonances in order to study its thermal excitation, either as a function of time or as a spectral density. Note that given the experimental configuration we measure the noise of the field probe polarization angle perpendicular to the analyzed component.

When the linearly polarized drive ($\theta_{\rm{L}}$) is kept on at a non-zero polarization angle relative to the ONF optical axis ($\theta_{\rm{F}}$), it can exert a torque on the nanofiber, which is proportional to the drive intensity, as Eqs. (\ref{torque}) and (\ref{eqmotion}) show. Depending on the relative angle $\Delta \theta$, the drive can increase or decrease the effective stiffness ($\kappa-\kappa_{\text{L}}(\Delta \theta)$) and damping ($\Gamma+\Gamma_{\text{L}}(\Delta \theta)$) of the rotational oscillator, leading to heating or cooling.

The driving process is not quite independent from the probing one. Even when the probe power is a hundred times smaller than the drive, we can see its effects. We notice that the cooling or heating conditions depend on the relative angle between the probe and the drive, suggesting that even small probe power can alter the birefringence of the ONF, being a key factor to determine the optical axis $\theta_{\rm{F}}$. This idea is strengthened by a slight dependence of the cooling performance relative to the probe power, suggesting that light-induced birefringence might be used for better engineering and control of the spin-optomechanical coupling in the future.

\section{Results}

\subsection*{Characterization of the thermal torsional noise}

We use two different methods to characterize the thermal noise of the system: the MB distribution of the rotational amplitude fluctuations in time and the resonance characteristics of the noise spectrum in frequency. Both approaches agree in the temperature changes within the experimental uncertainties.

The fluctuations of the angle $\theta_{\rm{F}}$ are transferred to the fluctuations of the probe light polarization angle, which we record as a voltage $V$. We measure the time series of the amplitude fluctuations of the noise around the mode resonance using the spectrum analyser. We keep the scan rate at zero, but widen the resolution bandwidth (RBW) ($\approx$50 Hz) to overlap the resonance whose FWHM is $\approx$3 Hz. We do not limit the video bandwidth in any way so it is larger than the resonance frequency ($\approx$190 kHz). The time series is usually 10 s with sampling at least every $5\times 10^{-4}$ s. 

We fit a MB distribution, Eq. (\ref{MB}), to a histogram of the  voltage $(V)$ of the time series. (The conversion between the angle and the voltage is  $\theta_F=\xi V$). The convergence of the fits to this particular nonlinear equation is complicated, but a change of variables that takes $\theta_{\rm{F}}^2=\Theta$ followed by natural logarithm converts Eq.~(\ref{MB}) into a straight line which is more robust to fit. The resulting slope carries the effective temperature information (${\rm{slope}}=-\kappa/2k_{\rm{B}}T_{\rm{eff}} \xi^2$). We show later the value of $\xi$ for our system based on the thermomechanical calibration of the ONF thermal fluctuations.  
 With such change of variable we have two parameters to fit the slope and the intercept.  The reduced $\chi ^2$ using the linear function to fit varies between 2 and 7. In order to bring those numbers to unity it is necessary to boost the statistical uncertainties by factors of about 2 in all cases. We check consistency of the results comparing the parameters to the average of the amplitudes squared Eq.~(\ref{PE-KE}), the results are all consistent (see Refs.~\cite{Kemiktarak2012,Laan2020} for this analysis). When data for the spectrum is available under the same experimental conditions, the  integral of the PSD in Eq.~(\ref{spdentheta}) gives consistent numbers with those extracted by the time series analysis.

\begin{figure}
\begin{center}
\includegraphics[width=7.0cm]{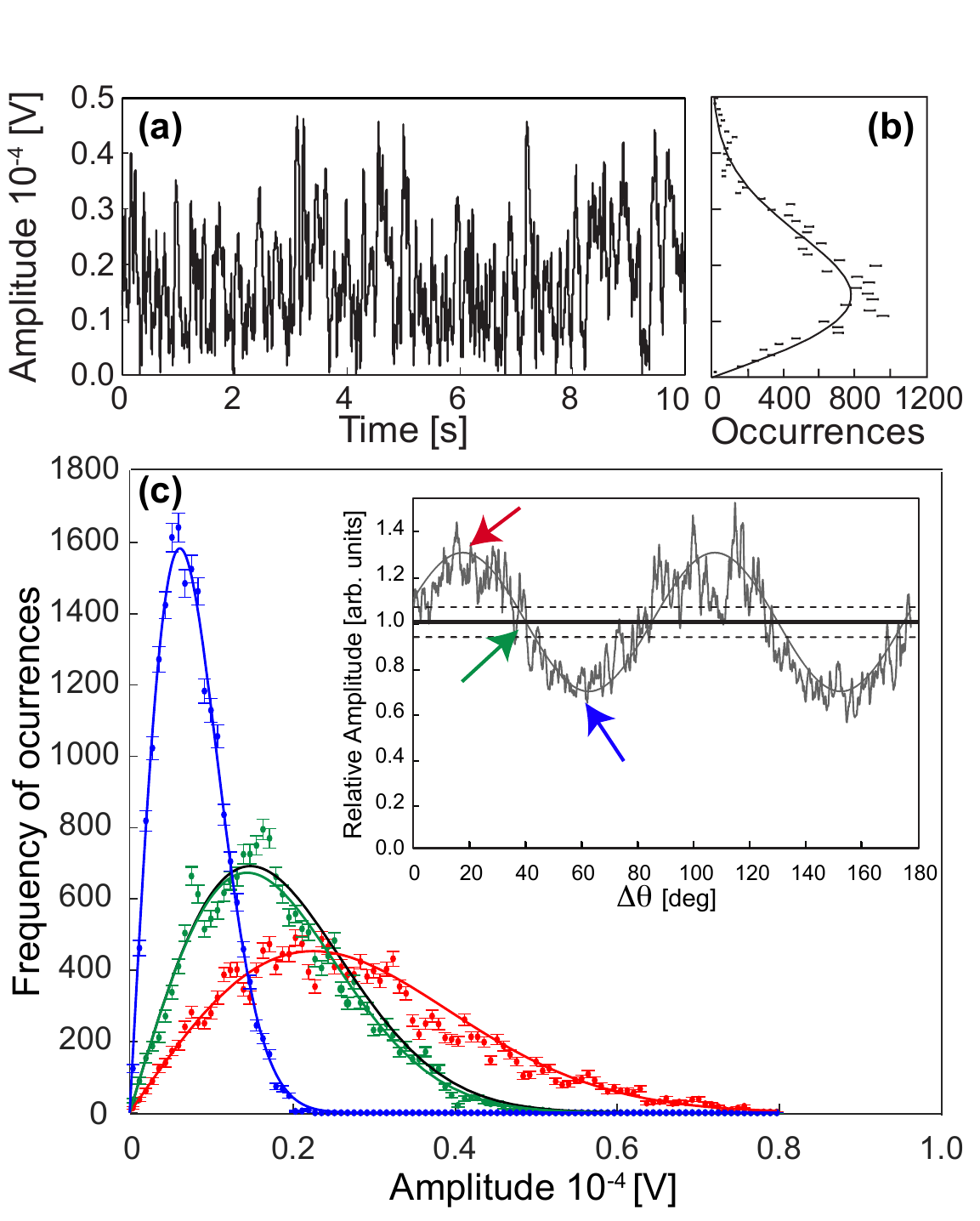}
\caption{(a) Time series of the amplitude fluctuations of the undriven first torsional mode. The plot shows 1 point every 20 from the 20,000 samples. The spectrum analyzer is fixed at the resonance frequency with RBW 50 Hz, and unlimited VBW. (b) Histogram of the data in (a) with 50 bins, showing a MB distribution. (c) MB distribution of the amplitude fluctuations of the first torsional mode for three different drive angles $\Delta \theta$. The continuous lines are fits to the MB distribution. The black line is the fit to the data from (a) with no drive agrees with the green data at the appropriate angle $\Delta \theta$. The inset shows the ratio of the noise amplitudes between the driven and undriven cases for the first torsional mode as a function of the drive angle $\Delta \theta$ under different experimental parameters. The solid gray line is the expected sinusoidal behaviour.  The dashed black lines bind the thermal noise level. The color arrows in the inset indicate the $\Delta \theta$ for the data in the main figure. See Table~\ref{table1} for the extracted temperature changes. 
}
\label{cooling}
\end{center}
\end{figure}

Figure \ref{cooling} shows the thermal distribution on the instantaneous oscillation amplitude of the voltage that comes from the polarization (proportional to $\theta_{\rm{F}}$ through $\xi$) sampled every 0.5 ms, under four different conditions.
The measurements of the thermal torsional noise at the mechanical resonance as a function of the drive polarization angle for the first torsional mode ($\nu_{T1}=$190.260 kHz) are done with a probe of 50 $\mu$W and a drive of 1 mW, under the digitazing conditions stated above.

Figure~\ref{cooling}(a) is a sample (one in every twenty points) of the amplitude fluctuations time series of the first torsional mode of the nanofiber with no drive. Fig.~\ref{cooling}(b) shows the histogram of the 20,000 points of the time series separated in 50 bins. Fig.~\ref{cooling}(c) shows MB distribution for three different time series taken with 1 mW drive, 50~$\mu$W probe and different drive angle $\Delta \theta$, (blue, green, and red traces on the data).  We fit Eq. (\ref{MB}) and its linear version to the data \cite{Kemiktarak2012} extracting the slope whose inverse is proportional to T$_{\rm{eff}}$.  Table \ref{table1} has the extracted numbers from the fits and their errors. The temperature from the no drive and the drive at 37.5$^{\rm{\circ}}$ are consistent showing the importance of $\Delta \theta$ in Eq.~(\ref{torque}).  The relative temperature decreases by a factor of 5.3 while the ratio of the average of the square to the square of the average remains around $4/\pi$ characteristic of  the MB distribution.

The sign of the optomechanical coupling has a period of $180^{\circ}$ with respect to the polarization angle of the drive (see Eq. (\ref{torque})), twice $\Delta \theta$ which corresponds to the half-waveplate angle of the experimental setup in the inset  of Fig.~\ref{cooling}. We observe such periodic modulation of the noise amplitude as a function of the drive polarization angle (see inset in Fig.~\ref{cooling}). The black trace is the fit to no drive case from Fig.~\ref{cooling}(b). Its MB distribution agrees with the fit of the $37.5^{\circ}$ case when the angle of the drive should cause no temperature change.
\begin{center}
\begin{table}
\begin{tabular}{  c c c c c } 
 \hline
 $\Delta \theta$, color & ---, black & 15$^{\circ}$, red & 37.5$^{\circ}$, green & 60$^{\circ}$, blue \\ 
 \hline
  Drive power [mW]& 0 & 1 &1 &1  \\ 
 
 1/Slope [$10^{-10}$V$^2$] & $4.14 \pm 0.06
 $&   $10.02 \pm 0.13$  & $4.000 \pm 0.005$ & $0.746\pm 0.005$ \\ 
   T$_{\Delta \theta}$/T$_{37.5^{\circ}}$ & $1.04 \pm 0.02$ & $2.56 \pm 0.03$ & $1 \pm 0.002$ & $0.187\pm 0.001$\\ 
   $\langle V^2\rangle/\langle V \rangle^2$ & $1.30 \pm 0.1$ & $1.29 \pm 0.01$ & $1.26 \pm 0.01$ & $1.26\pm 0.01$\\ 
   \hline
\end{tabular}
   \caption{Parameters extracted from the fits to MB distributions for the plots in Fig. \ref{cooling}.}
   \label{table1}
   \end{table}
\end{center}

 \begin{figure}
\begin{center}
\includegraphics[width=6.9cm]{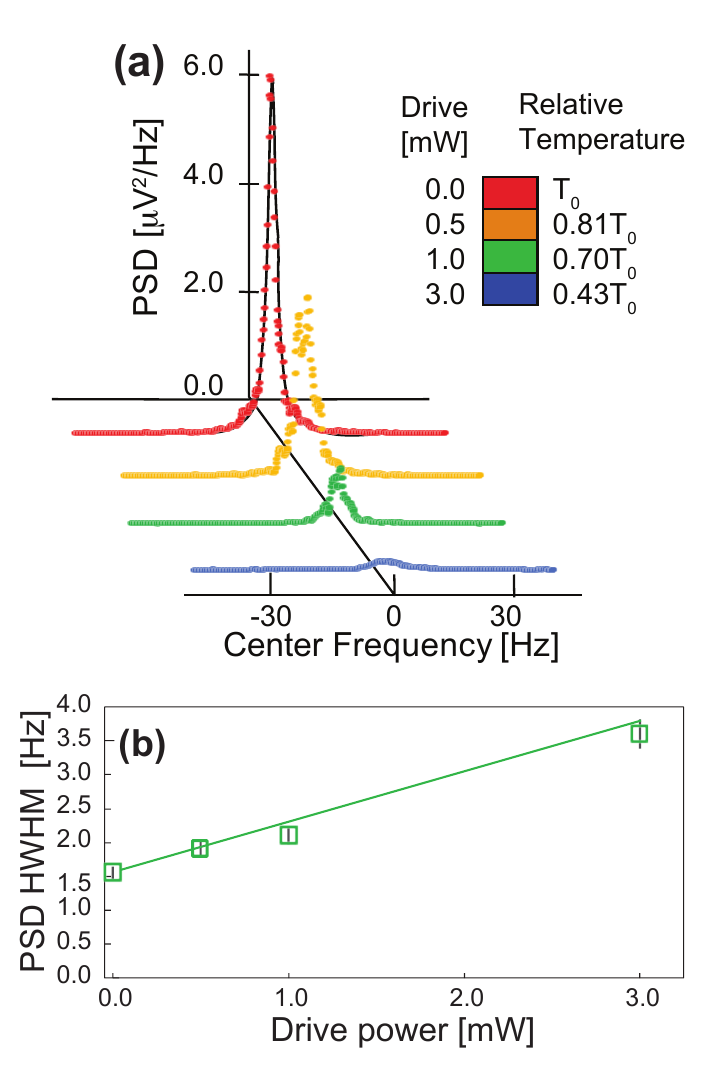}
\caption{Power Spectral Density study as a function of drive power of the first torsional mode.
(a) Evolution of the PSD as the drive power increases, with probe 50 $\mu$W. The spectra are plotted with their center frequency aligned for ease of presentation. The no drive resonance is at 189,664 Hz. The color code typifies the relative cooling with respect to no drive $T_{\rm{0}}$. The continuous black line on the 0 drive (red) spectrum is a fit to the PSD from Eq.~(\ref{spdentheta}). (b) Change in the PSD HWHM ($\Gamma_{\rm{eff}}$) as a function of the drive power for the amplitude spectra shown on part (a) with a linear trend as a continuous green line.
}
\label{power}
\end{center}
\end{figure}

The second method we use to characterize cooling is from the PSD that we get from the amplitude spectral density with a spectrum analyzer, as shown in Fig. \ref{power}. We scan 250 Hz around the resonance with RBW 1 Hz and VBW 1 Hz. The expected functional form is Eq.~(\ref{spdentheta}). We fit the data to that function using as parameters the amplitude, HWHM $(\Gamma_{\rm{eff}})$, and center frequency $(\omega_{\rm{eff}})$. We do not have a statistical uncertainty for the data, but use a fixed percentage (between 2 an 5\%) to obtain reduced $\chi^2 \approx 1$. The no drive (red data points) fit with the continuous black line is an example of such a fit. The integral of the power spectral density gives us a characterization of the temperature in Eq.~(\ref{PE-KE}) and the number is consistent with those obtained with the time series method. Fig. \ref{power} (a) shows measurements of the power spectral density for different drive power. As the optical power increases, and the cooling improves, the PSD reduces its amplitude while broadening, The relative scale of the temperatures (based on the integral under the curve) for different drive power is color-coded as function of the no drive temperature $T_{\rm{0}}$. Fig. \ref{power} (b) shows how the PSD HWHM follows a linear trend, which agrees with Eq.~(\ref{SM.Eq.Gamma2}). We found that at higher powers ($>$10 mW) the mechanical stability of the ONF appears compromised. We observe large fluctuations on the Rayleigh light scattered out of the nanofiber, suggesting that vibrational (non-torsional) modes are excited.

\subsection*{Frequency Shift of the Resonance}

Optomechanical cooling implies changes in the linewidth and a frequency shift of the mechanical resonance, as Eq. (\ref{spdentheta}) shows. The measurements of the optomechanicaly-induced frequency shifts is complicated by the fact that at some optical power the ONF starts to radiate thermal energy. Its thermal equilibrium approximately follows a $T^3$ dependence, since the radiation from a thin object differs from a 3-dimensional blackbody \cite{wuttke2013prl}. The reached equilibrium temperature modifies the ONF mechanical properties and its resonant frequency. Our measurements at higher drive powers show qualitative agreement with such process \cite{fenton2016}, but the effects are not decoupled enough to extract the appropriate optomechanicaly-induced frequency shifts. Nonetheless, we see changes of $\sim$10 Hz of the resonant frequency at constant lasers power by rotating the drive polarization angle through heating and cooling conditions, allowing us to estimate the magnitude of the optomechanically-induced frequency shift.
 \begin{figure}
\begin{center}
\includegraphics[width=0.5
\textwidth]{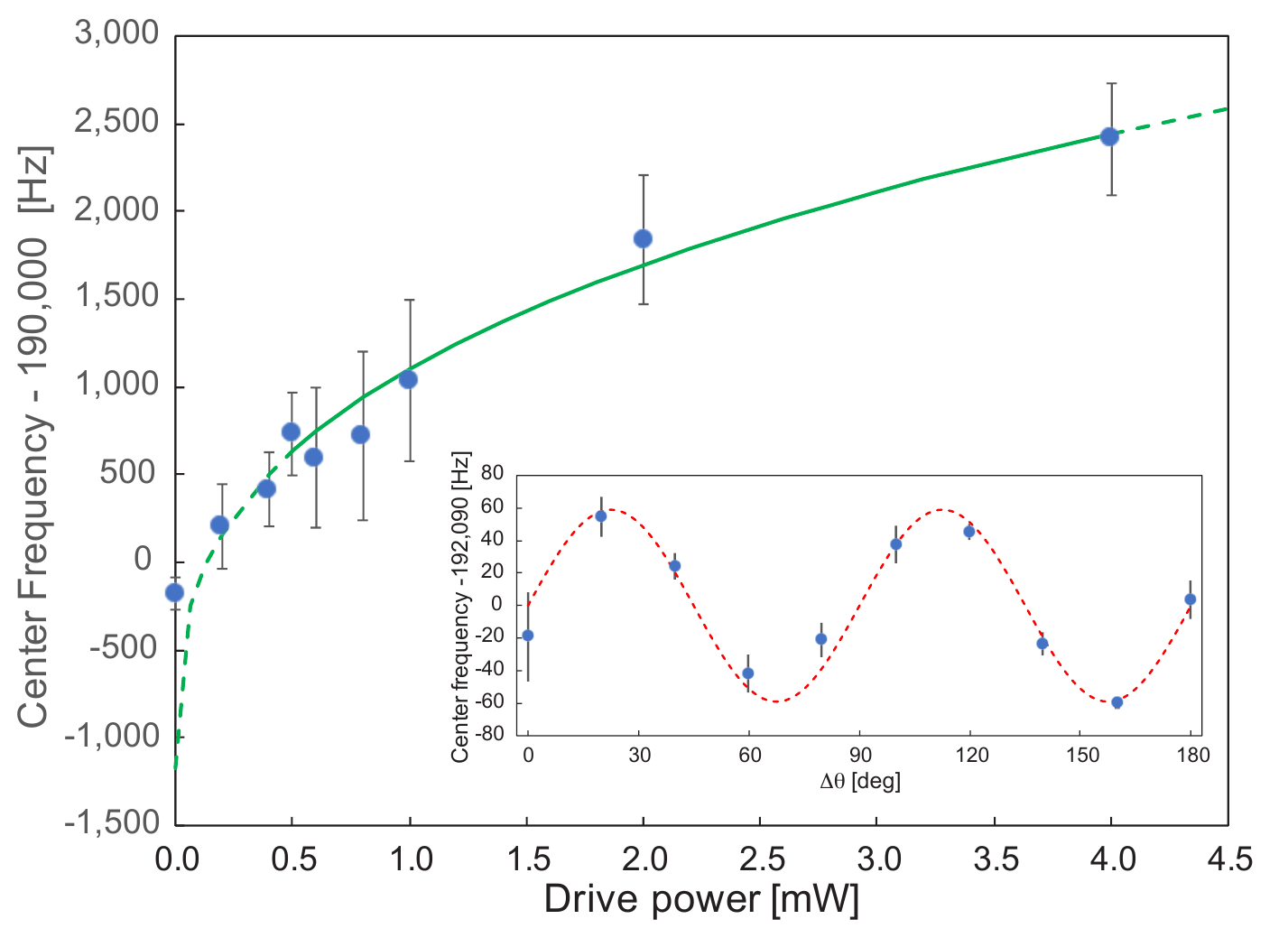}
\caption{Frequency shift of the resonance of the first torsional mode as a function of the drive power (RBW 1 Hz, VBW 1 Hz). The green solid line is a fit to an offset, a linear term and cubic root starting at the 0.5 mW point. The green dashed lines is the extrapolations from the fit.  The inset shows the frequency shift of the
resonance as a function of drive polarization angle for a fixed drive power of 10 mW. The red dashed line is a fit to a sinusoidal function.} 
\vspace{-0.7cm}
\label{shift}
\end{center}
\end{figure}

Figure \ref{shift} shows the shift of the mechanical resonance as a function of laser power. The error bars come from the dispersion of six independent sets of measurements that have been averaged (some of them from the conditions of Fig.~\ref{cooling}), with a center frequency error of approximately 1Hz each.  The no drive point is quite stable. The green continuous line is a fit to the points to $y=ax^{1/3}+b$, with y the center frequency and x the power in mW, between 0.5 mW and 4.0 mW. The cubic root dependence, already pointed in the work of Fenton et al. \cite{fenton2016} is visible, but if we look for linear dependence, add a term $cx$ on the green fit, the slope $c$ is consistent with zero with an error of $\pm 40$ Hz/mW. The error is larger than the shifts we observe at a fixed drive by changing the angle between the probe and the drive polarizations, which are of the order of 10 Hz.

The inset in Fig. \ref{shift} shows the frequency shift (amplitude  about 40 Hz) with 10 mW drive as a function of the angle of the drive similar to the noise amplitude change (see inset in Fig.~\ref{cooling}). The result is the average of the oscillations around the mean of three consecutive measurements, the errors are the standard deviation of the mean. The red line dashed follows the expected sinusoidal behavior.
The measurement is sensitive to the small power changes
as a function of the drive angle, which thermally induce
changes of the resonant frequency, preventing us from quantitative assessment of the exact shifts. However, we do observe how they increase when the optomechanical drive polarization angles is at the heating position, and decrease when it is at the cooling position.

The ratio between the optically induced and bare torsional spring constant is related to the frequency shift as $\delta=1-(\omega_{\rm{eff}}/\omega_0)^2=\kappa_{\rm{L}}/2\kappa$. Then, from Eqs.~(\ref{SM.Eq.T}) and (\ref{SM.Eq.Gamma2}) we can estimate an effective temperature in a different way
\begin{equation}
T_{\rm{eff}}\sim T \frac{1}{1+\delta Q},
\end{equation}
where we explicitly see that the optomechanical cooling depends on the quality factor of the mechanical resonator and the light-induced frequency shift. For our particular experiment, we report a $Q=6\times 10^4$ and we observe a frequency shift of about 10 Hz for 2 mW drive that corresponds to $\delta\approx 10^{-4}$. Then, we estimate a temperature reduction of $
T_{\rm{eff}}\sim T/7$, which agrees with the magnitude of the observed cooling (see Table~\ref{table1}).

\subsection*{Rotation Amplitude Estimate}

There are two distinct types of rotations that take place in our system: The mechanical rotations of the ONF and the polarization rotations of the probe light, used to measure the mechanical ones. The polarization rotation is a proxy for the mechanical rotations and, although they are linearly proportional to each other, they are generally not of the same amplitude. Here we estimate the rotation amplitude of both the mechanical rotation and the light polarization.

One can thermomechanicaly calibrate the system and estimate the ONF torsional angle from its mechanical and thermal properties at room temperature ($T=300K$). Eq. (\ref{PE-KE}) states that in the absence of an external drive $\langle|\theta_{\rm{F}}|^{2}\rangle=k_{\rm{B}}T/\kappa$. The torsional spring constant can be determined from $\kappa=\omega_0^2 I$, where the moment of inertia (simplifying to a rod with no tapers) is $I=\pi \rho l r^4/2$. The moment of inertia is typically three to five orders of magnitude larger than in nanorod experiments (see Ref.~\cite{Kuhn17Optica}), given the macroscopic length of the ONF. Considering that the density of fused silica is $\rho=2.2 \times 10^3$ kg/m$^3$, the radius and length of the nanofiber waist are $r=390\times10^{-9}$ m and $l=5 \times 10^{-3}$ m, and the rotational angular frequency is
$\omega_0=2 \pi \times 1.9\times 10^5$ rad/s,  we get that $k_BT/\kappa=7.3\times 10^{-9}$. Compared to the area of the measured power noise spectrum, consistent with the values extracted from the time series fits, with no drive of the fundamental mode (see Fig.~\ref{cooling}b) of $\approx 5\times 10^{-10}$~V$^2$, we obtain a voltage-to-angle conversion factor of $\approx$ $\xi=3.8$ rad/V.  We can then estimate an order of magnitude of the ONF torsional angle from the typical 5 $\mu$V signal of $\langle|\theta_{\rm{F}}|\rangle\sim 20~\mu$rad.

The optical power oscillating at the frequency of the first torsional mode is the result of the beat between the probe field $E_{0}$ and the orthogonal component of the electric field $\delta E_{y}$. Assuming the polarization rotation angle $\phi$ to be small enough, then $\delta E_y=\phi E_{0}$. When measuring the thermally excited torsional modes with a laser probe power of $50~\mu$W, the voltage amplitude at the output of the Balance Photo Diode is typically $V_{\rm{AC}}\approx 5~\mu$V. Given the gain of the RF output of $ 10^4$ V/A and a responsivity of 0.5 A/W  this corresponds to a fluctuating optical power of $P_-\approx25$ pW, while the optical power on each detector is $P_{(H,V)}=25 \mu$W. This gives a rotation angle estimate of $\phi=E_0\delta E_y/|E_0|^2=i_-/4i_{(H,V)}=P_-/4P_{(H,V)}=0.25 \times10^{-6}$ rad. The voltage-to-angle conversion factor is then the ratio of this rotation angle to the original voltage $V_{\rm{AC}}$, giving a signal of polarization rotation of $\approx$ 0.05 rad/V which is smaller than the mechanical estimate.

Due to the reflection symmetry of the torsional modes relative to a cross-sectional plane at the center of the ONF, most of the light polarization rotation experienced by the probe up to the middle of the nanofiber becomes undone during the other half of the propagation. Only asymmetries in the system allow to observe a net polarization rotation. So a polarization rotation of $\phi\sim ~ 1 \mu$rad is a lower bound to the actual mechanical rotation of the ONF and is consistent with the estimate above.

\subsection*{Feedback observations}
\begin{figure}
\begin{center}
\includegraphics[width=0.5
\textwidth]{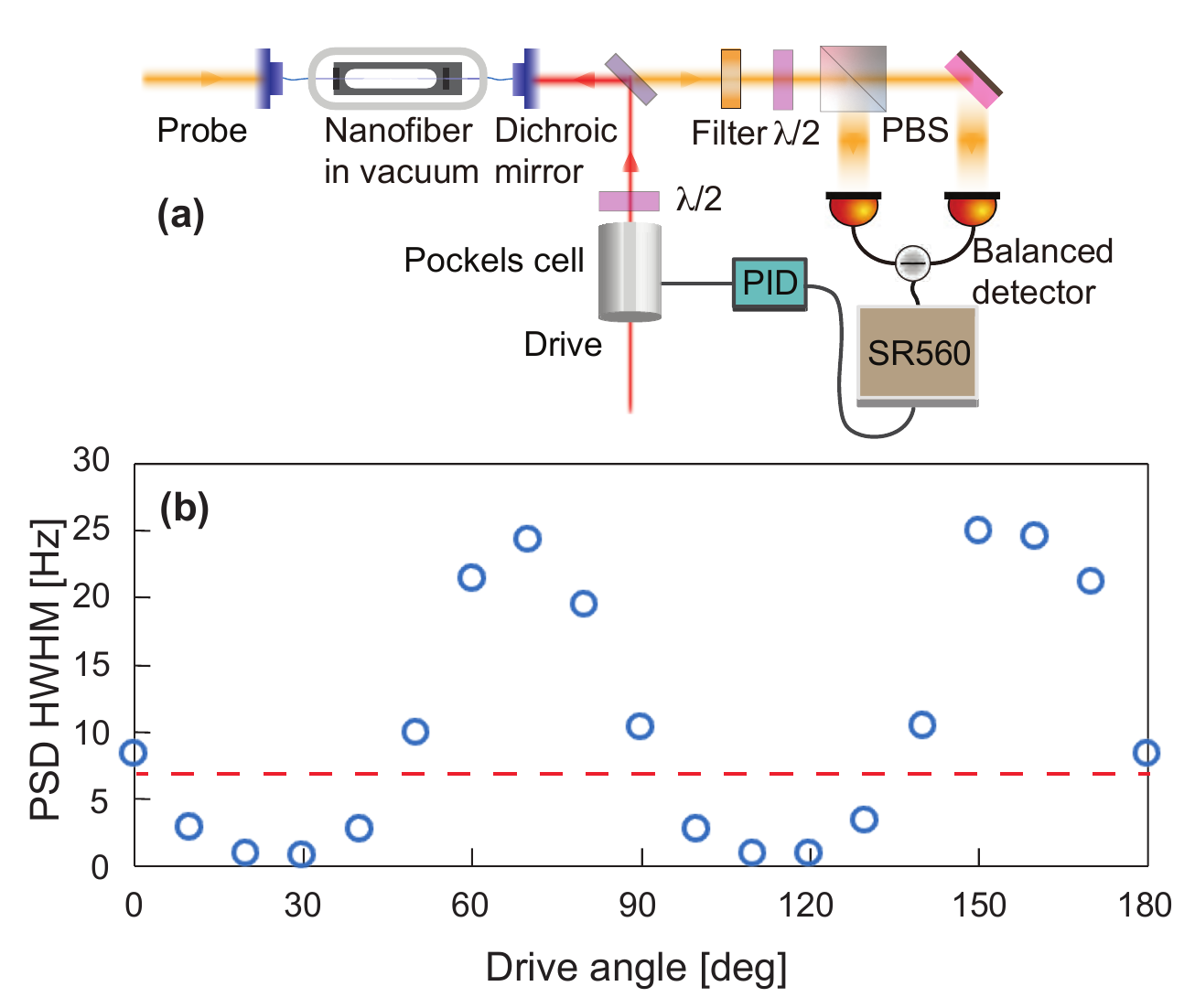}
\caption{(a) Feedback schematic. (b) Power HWHM of the PSD as a function of drive angle, the fit errors are smaller than the circles. The RBW and VBW for the measurements of the spectra are 1 Hz with drive at 1 mW and probe at 60 $\mu$W.  The dashed line corresponds to the case without drive and sets the reference for an effective room temperature.} 
\label{fb}
\end{center}
\end{figure}

Active feedback can dramatically improve the performance of optomechanical cooling. A proof-of-principle test demonstrates that such strategies are realizable in the presented platform. Figure \ref{fb}(a) shows the schematic of the feedback setup. The nanofiber in the tests has low intrinsic birefringence (characterized by $\eta$ in Eq.~(\ref{torque0})) and the ONF does not show any measurable optomechanical cooling under normal circumstances. This is compounded with the degradation of its mechanical quality, manifested by a larger HWHM of about 7 Hz and a decreased Q. However, when the active feedback is on, the spectral width increases (decreases) and by measuring the areas of the PSD around resonance we infer a reduction (increment) of the temperature of almost a factor of four (see Fig.~\ref{fb}(b)). A simple feedback, of just modulating the amplitude of the drive at the resonance frequencies and fixing the drive polarization angle, allows a significant optomechanical effect. This opens the avenue to further investigations.

\section*{Discussion and outlook} 
We observe that all higher order torsional modes decrease their amplitudes when the fundamental mode decreases. This demonstrates that the optomechanical coupling is broadband, affecting all mechanical modes simultaneously \cite{Kemiktarak2012}. Moreover, the scheme does not need an optical cavity.
This characteristic is quite different from typical optomechanical cooling that relies on an optical cavity with a narrow bandwidth driven near optical resonance \cite{aspermeyer14}.

The PSD around the resonant frequency of a torsional mode can increase with external mechanical vibrations of much lower frequencies. This may be due to the mechanical coupling between the torsional and string modes of the ONF. Future studies could benefit from measuring and characterizing the string modes to further understand the effects of such mechanical cross-coupling under optomechanical cooling of the torsional modes.

Looking back at Eqs.~(\ref{torque},\ref{torque0}) there is a second sine function associated with the difference of the two indices of refraction. If the probe affects those indices through a Kerr effect, then a second modulation appears in the torque creating a richer phase space to explore. We have mapped the dependence of the heating and cooling as a function of the two angles, the one for the probe and the one for the drive. The dependence of the optomechanical coupling on the probe laser power and polarization angle indicates an optical non-linearity and a possible cross-coupling between distinct optical beams. The optomechanical platform we present might have a rich range of phenomenon to explore, and broad possibilities for feedback cooling. On the other hand, torsional modes can be selectively driven by a periodic modulation of the drive power \cite{fenton18}. We observe hysteresis in the amplitude of the oscillation when sweeping the modulation frequency of the drive, heralding many more studies of torsional optomechanical coupling with non-linear dynamics that may include bistability. 

Atomic optical dipole traps in nanofibers can benefit from the shown mechanisms to quiet torsional modes, specially when the mechanical frequencies are comparable to the trapping frequencies \cite{solano17c}. One can also envision the engineering of suspended optical waveguides with transverse long arms to create torsional pendula \cite{kim17}. 

In order to observe quantum effects \cite{Stickler2021}, the mechanical system has to be at a temperature $T<(f\times Q)h/k_B$ \cite{Chakram2014}.
Although in our case this corresponds to $T<0.3~K$, it should be possible to design and fabricate suspended optical waveguides for this purpose. Aiming for higher-frequency torsional modes and higher Q-factor might help reaching the quantum limit at experimentally achievable temperatures. Lower temperatures can be achieved provided a larger optomechanical coupling. To increase the optomechanical coupling one could fabricate waveguides with a transverse asymmetry or high birefringence. 

Our results show cooling of the torsional mode of a macroscopically extended object via angular momentum exchange. The measured temperature reduction is far from that of state-of-the-art optomechanical systems. However, the demonstration was realized on a versatile system that was not designed or optimize with this goal, suggesting that there is plenty of room for improvements on the theoretical and experimental aspects of this novel phenomenon.

\section*{Summary} We have demonstrated cooling of a rotational degree of freedom using the optomechanical coupling between the torsional modes of an ONF and the polarization of light propagating through it. In particular, we measured a factor of five reduction of the effective temperature of the torsional motion without the need for an optical resonator or feedback. This provides the first demonstration of optomechanical cooling by light-matter angular momentum exchange of a macroscopic torsional mode. 
We expect that a better control of the presented effect, through optimized design and fabrication of the torsional waveguide combined with optical feedback, will allow for noise reduction in torsional pendula for precision measurements with macroscopic objects.

\section*{Funding}
This work was supported by the National Key R\&D Program of China (Grant No. 2017YFA0304203), the Natural Science Foundation of China (Nos. 6210031464, 61675120, 12034012, 61875110), the Natural Science Foundation of China Project for Excellent Research Team (No. 61121064), Shanxi '1331 Project' Key Subjects Construction, PCSIRT (No. IRT\_17R70) and 111 project (Grant No. D18001) from China, CONICYT-PAI 77190033 and FONDECYT 11200192 from Chile.

\section*{Acknowledgments}
We are grateful to Howard J. Carmichael, John Lawall, and Nergis Mavalvala  for useful discussions. Special thanks to J. Lawall for his constant feedback during the elaboration of the manuscript. L. A. O. thanks the hospitality of the Institute of Laser Spectroscopy, Shanxi University, China, where this experiment  has been performed. 

%

\bibliography{refs}

\end{document}